\documentclass[
twocolumn,
superscriptaddress,
bibnotes,
amsmath,amssymb,
aps,
pra,
]{revtex4-1}

\usepackage{amsmath,amssymb,epsfig}
\usepackage{graphicx}
\usepackage{dcolumn}
\usepackage{bm}
\usepackage[normalem]{ulem}
\newcommand{\braket}[1]{\left\langle #1 \right\rangle}
\newcommand{\bra}[1]{\left \langle #1 \right|}
\newcommand{\ket}[1]{\left| #1 \right \rangle}

\usepackage{bm, color}
\definecolor{ryangreen}{rgb}{0.0,0.8,0.2}

\usepackage{hyperref}
\hypersetup{
    colorlinks=true,
    linkcolor=blue,
    filecolor=magenta,      
    urlcolor=blue,
    citecolor=blue,
    }

\begin{document}

\preprint{APS/123-QED}

\title{BCS-BEC crossover of the strongly interacting $^6$Li-$^{40}$K mixture}

\author{Stefano Gandolfi}
\affiliation{Theoretical Division, Los Alamos National Laboratory, Los Alamos, New Mexico 87545, USA}
\author{Ryan Curry}
\affiliation{Department of Physics, University of Guelph, Guelph, Ontario N1G 2W1, Canada}
\affiliation{Theoretical Division, Los Alamos National Laboratory, Los Alamos, New Mexico 87545, USA}
\author{Alexandros Gezerlis}
\affiliation{Department of Physics, University of Guelph, Guelph, Ontario N1G 2W1, Canada}


\begin{abstract}
We present quantum Monte Carlo calculations of the properties of a
two-component mass imbalanced Fermi gas,
corresponding to the $^6$Li-$^{40}$K mixture.  We compute the equation of
state of the unpolarized system as a function of the scattering length
with particular attention paid to the unitary limit, where the effect
of the effective range of the interaction is explored. In order to investigate differences from the equal-mass case we also
compute the pair-distribution function and the momentum distribution
over a range of interaction strengths, which can provide information about
the structure of the system.  Finally, we compute the heavy-light
quasiparticle spectrum for the full crossover regime.  Our theoretical predictions,
based on quantum Monte Carlo calculations, should inform future theoretical studies and can be tested by experiments
with ultracold fermionic gases.
\end{abstract}

\maketitle


\section{Introduction}
In recent years much work has been done to study ultracold
Fermi gases, both experimentally and theoretically (for a complete
review see for example Ref.~\cite{Giorgini_Pitaevskii_Stringari_2008}).  Through the use of a Feshbach resonance it is possible to tune the interaction between fermions as
desired~\cite{Chin_Grimm_Julienne_etal_2010}, and the amazing experimental
control on these systems gives access to beautiful experiments
in which Fermi gases can be studied under a range of conditions.  It is
possible to probe the entire BCS-BEC crossover \cite{Strinati_Pieri_Ropke_etal_2018}, where the system can be tuned from a BCS state where fermions are weakly interacting, to
the unitary regime where the two-body scattering length is infinite,
to a Bose-Einstein condensate (BEC) of bosons.  When the population
imbalance changes, the system can eventually exhibit different
phases~\cite{Zwierlein_Schirotzek_Schunck_etal_2006,Partridge_Li_Kamar_etal_2006} where superfluid and normal
phases may coexist or phase separate~\cite{Pilati_Giorgini_2008,Lobo_Recati_Giorgini_etal_2006} or
other intriguing more exotic phases like the Larkin–Ovchinnikov–Fulde–Ferrell (LOFF) phase or $p$-wave superfluidity
could appear~\cite{Bulgac_Forbes_Schwenk_2006,Bulgac_Forbes_2008}.

Considerable experimental effort has been spent on the investigation of trapped mass imbalanced two-component Fermi gases. Early work explored mainly systems consisting of a mixture of $^6$Li and $^{40}$K atoms ~\cite{Wille_Spiegelhalder_Kerner_etal_2008,Taglieber_Voigt_Aoki_etal_2008,Spiegelhalder_Trenkwalder_Naik_etal_2010,Tiecke_Goosen_Ludewig_etal_2010,Voigt_Taglieber_Costa_etal_2009,Trenkwalder_Kohstall_Zaccanti_etal_2011}, and in recent years experimental probes have been conducted for a wide variety of mass-imbalanced systems \cite{Khramov_Hansen_Dowd_etal_2014,Ravensbergen_Corre_Soave_etal_2018, Ravensbergen_Soave_Corre_etal_2020, Neri_Ciamei_Simonelli_etal_2020, Ciamei_Finelli_Cosco_etal_2022, Green_Li_SeeToh_etal_2020, Schafer_Haruna_Takahashi_2023} with various mass ratios.  There has also been continued interest in probes of the $^6$Li-$^{40}$K mixture, such as investigations into dimer stability \cite{Jag_Cetina_Lous_etal_2016}.

The properties of the mass-imbalanced system can be very
different from the equal-mass case; for example, with a majority light
population the Chandrasekhar-Clogston limit is very small and close
to zero~\cite{Gezerlis_Gandolfi_Schmidt_etal_2009}. Then, at unitarity, a two-component
Fermi gas with unequal masses may exhibit very different properties with
respect to the equal-mass case~\cite{Gubbels_Baarsma_Stoof_2009,Baarsma_Gubbels_Stoof_2010}, even at the mean field level~\cite{Baranov_Lobo_Shlyapnikov_2008, Wang_Che_Zhang_etal_2017}.  Systems of mass-imbalanced fermions also show new features in the few-body sector.  The binding energy
of an impurity was calculated in Ref.~\cite{Combescot_Recati_Lobo_etal_2007}; however,
finite systems with few-heavy atoms that interact with a single light one through
a two-body potential with infinite scattering length have very intriguing
nonuniversal properties~\cite{Blume_Daily_2010,Blume_Daily_2010a,Castin_Mora_Pricoupenko_2010},
and this could effectively modify the many-body structure of the state
with polarization, supporting new phases that do not show in the equal
mass case \cite{Braun_Drut_Jahn_etal_2014, Roscher_Braun_Drut_2015,Wang_Che_Zhang_etal_2017, Pini_Pieri_Grimm_etal_2021}.
It has also been shown that for very large mass ratios the ground state of 
the unpolarized system might be solid~\cite{Petrov_Astrakharchik_Papoular_etal_2007, Osychenko_Astrakharchik_Mazzanti_etal_2012}. In addition, there is rich physics to explore in mass-imbalanced systems existing in lower dimensions, such as novel types of pairing \cite{Rammelmuller_Drut_Braun_2020} and quarteting \cite{Liu_Wang_Cui_2023}. 

Since experiments are possible at very low
temperatures, of the order of fractions of $T_F$, the zero-temperature effects can directly be compared to or extrapolated from
experiments. Several $T=0$ predictions for ultracold Fermi gases were made in the
past by means of quantum Monte Carlo (QMC) techniques \cite{Carlson_Chang_Pandharipande_etal_2003,Chang_Pandharipande_Carlson_etal_2004,Carlson_Reddy_2005,Carlson_Reddy_2008,Astrakharchik_Boronat_Casulleras_etal_2004, Gezerlis_Gandolfi_Schmidt_etal_2009, Carlson_Gandolfi_Schmidt_etal_2011, Braun_Drut_Roscher_2015}, and were later confirmed by
experiments for both the equal-mass \cite{Ku_Sommer_Cheuk_etal_2012} and unequal-mass \cite{Kohstall_Zaccanti_Jag_etal_2012} cases.
The variational Monte Carlo (VMC) and the diffusion Monte Carlo (DMC) 
methods allow one to accurately solve for the ground state of strongly interacting many body systems \cite{footnote}. The QMC family of approaches has been widely used in the study of diverse physical systems such as condensed matter, quantum chemistry, nuclear physics, and cold atomic physics \cite{Ceperley_1996a, Foulkes_Mitas_Needs_etal_2001, Kolorenc_Mitas_2011, Roggero_Mukherjee_Pederiva_2013, Carlson_Gandolfi_Pederiva_etal_2015, Schonenberg_Conduit_2017, Gandolfi_Carlson_Roggero_etal_2018, Curry_Lynn_Schmidt_etal_2023}. Previous studies explored the properties of strongly interacting fermions at unitarity~\cite{Gandolfi_Schmidt_Carlson_2011} however, the theoretical study of the unequal mass system has not received effort similar to the equal-mass case.

In this paper we focus on a Fermi-Fermi mixture with a mass ratio
$m_h/m_l$=6.5 corresponding to the $^6$Li-$^{40}$K mixture, and we
study the BCS-BEC crossover by employing QMC techniques; in particular
in this work we compute the properties of the unpolarized system with an equal
number of heavy and light atoms.  We first compute the equation of state
that is crucial to fit the density functionals used to compute the properties of
trapped larger systems~\cite{Forbes_Gandolfi_Gezerlis_2011,Bausmerth_Recati_Stringari_2009,Forbes_Gandolfi_Gezerlis_2012}. The preliminary results of these calculations were presented in \cite{Gandolfi_2014}, although we have considerably expanded upon that early work, addressing a number of physical observables. We compute the energy per particle as a function of the scattering length of the two-body interaction by paying particular attention to the effect of
the finite effective range of the potential.  We also compute
the pair-distribution function between heavy and light particles as well as
the momentum distribution.  These quantities are useful for extracting the
contact parameter that can be measured in experiments, similar to the equal-mass case~\cite{Tan_2008,Tan_2008b,Tan_2008c}.  We then simulate the
system with very small polarization in order to study the quasiparticle dispersion as a
function of the interaction. These calculations serve to carefully explore the properties of the $^6$Li-$^{40}$K system and provide stringent theoretical predictions for future experiments.

\section{Model}
We model our system by considering point-like interacting particles 
with the following Hamiltonian:
\begin{equation}
H=\frac{-\hbar^2}{2m_l}\sum_{i=1}^{N_l}\nabla_i^2
+\frac{-\hbar^2}{2 m_h}\sum_{j^{\prime}=1}^{N_h}\nabla_{j^\prime}^2
+\sum_{i<j^{\prime}}v(r_{ij^{\prime}}) \,,
\end{equation}
where the sum $i$ is over $N_l$ light particles with mass $m_l$ and the $j'$ sum is over $N_h$ heavy particles with mass $m_h$. $v(r_{ij^{\prime}})$ is an $s-$wave short-range interaction acting only between
fermions of different species.
The form of $v(r)$ we use is the P\"oschl-Teller
already employed in several previous
QMC calculations~\cite{Carlson_Chang_Pandharipande_etal_2003,Chang_Pandharipande_Carlson_etal_2004,Carlson_Reddy_2005,Gezerlis_Gandolfi_Schmidt_etal_2009,Morris_LopezRios_Needs_2010,Gandolfi_Schmidt_Carlson_2011,Forbes_Gandolfi_Gezerlis_2011,Forbes_Gandolfi_Gezerlis_2012}:
\begin{equation}
v(r)=-v_0\frac{\hbar^2}{m_r}\frac{\mu^2}{\cosh^2(\mu r)} \,,
\end{equation}
where $m_r$ is the reduced mass, and the parameters $\mu$ and $v_0$ are tuned in order to reproduce the effective range $r_e$ and the scattering length $a$
of the interaction. For this potential, the unitary limit corresponding
to the zero-energy ground-state between two particles is achieved with $v_0=1$
and $r_{e}=2/\mu$.  We consider a system in the dilute limit, meaning
that the interparticle distance $r_0 \gg r_{e}$, where
$r_0=(9\pi/4)^{1/3}/k_F$ and $k_F$ is the Fermi momentum of the system.  Most of the results presented in this work
were obtained by fixing $r_{e} k_F\approx 0.03$, but in several cases we
considered different values to check effects due to the effective range
of the potential.

The ground state of the system is solved through the use of QMC
techniques, in particular VMC and DMC.
The many-body wave function ~\cite{Carlson_Chang_Pandharipande_etal_2003,Chang_Pandharipande_Carlson_etal_2004} is given by 
\begin{equation}
\label{eq:wf}
\Psi_T =  \left[\prod_{i<j'} f(r_{ij'})\right] \Phi_{\rm BCS} \,.
\end{equation}
The Jastrow function $f(r)$ acts between particles with different masses,
and it is obtained by solving the equation
\begin{equation}
\label{eq:jas}
-\frac{\hbar^2}{2m_r}\nabla^2f(r)+v(r)f(r)=\lambda f(r) \,.
\end{equation}
The parameter $\lambda$ is obtained by imposing the boundary condition
$f(r>d_0)=1$, where the healing distance $d_0$ is a variational parameter.
The correct boundary conditions of the wave function are guaranteed
by constraining $d_0\le L/2$, where $L$ is the size of the simulation
box, but we find that a typical good choice is $d_0\approx L/10$.
By solving Eq.~(\ref{eq:jas}) we also ensure the correct behavior of the
wave function at small distances so that the Jastrow function $f$ is
defined to have $\partial f/\partial r =0 $ at the origin.  This condition
must be satisfied in order to have a finite kinetic energy in the origin
when computing $\nabla^2 f=f''(r)+2f'(r)/r$.
This condition is necessary to avoid spurious contributions to the energy
computed using QMC.

The antisymmetric part $\Phi_{\rm BCS}$ is a particle-projected BCS
wave function including pairing correlations (see the Appendix of
Ref.~\cite{Gandolfi_Illarionov_Pederiva_etal_2009}).  It is given by 
\begin{equation} 
\Phi_{\rm BCS} = {{\cal A}}[\phi(r_{11'}) \phi(r_{22'}) ... \phi(r_{N_lN_h})] \,.
\end{equation} 
The operator ${\cal A}$ antisymmetrizes like particles,
the unprimed coordinates are for light particles, the primed ones are
for heavy particles, and $N_l=N_h=N/2$ for the unpolarized case. The pairing function is expressed as
\begin{eqnarray} 
\label{eq:pairfn} 
\phi({\bm r}) &=&\beta (r) + \sum_{\bm{n},I\leq I_C} a(k_{\bm{n}}^2) \exp [ i \bm k_{\bm{n}} \cdot \bm r]~, 
\nonumber\\
\beta(r)&=& \tilde{\beta}(r)+\tilde{\beta}(L-r)-2 \tilde{\beta}(L/2)~,
\nonumber\\ 
\tilde{\beta} (r) &=& [ 1 + \gamma b  r ]\ [ 1 - \exp ( -d b r )] \frac{\exp ( - b r )}{dbr}~.  
\end{eqnarray} 
The function $\beta(r)$ has a range of $L/2$, and the value of $\gamma$ is
chosen such that $\beta(r)$ has zero slope at the origin. The sum in the pairing function $\phi(\bm{r})$ is over $\bm{k}_{\bm{n}} = \frac{2\pi}{L}(n_x,n_y,n_z)$, which imposes a shell structure according to $I=n_x^2 + n_y^2 + n_z^2$ that we cutoff at some value $I_C$.
Note that if $\beta(r)=0$, and $a(k_{n}^2)=0$ for $|\bm{k}_n| > k_F$, $\Phi_{\rm BCS}=\Phi_{FG}$,
where the latter is a Slater determinant wave function describing the noninteracting
Fermi gas in the normal phase. As $\Phi_{FG}$ and $\Phi_{\rm BCS}$
are orthogonal, we can study the Fermi gases both in the superfluid
and in the normal phase~\cite{Lobo_Recati_Giorgini_etal_2006,Pilati_Giorgini_2008}. The pairing wave
function used here contains 12 free parameters that have been optimized
by minimizing the energy of the system using VMC and following the
strategy of Ref.~\cite{Sorella_2001}.  In the case of strong interactions,
when $1/ak_F\geq 2$, we found that using the pairing function as in
Ref.~\cite{Astrakharchik_Boronat_Casulleras_etal_2004}, gives lower energies. Practically this involves finding the solution to the two-body Schr\"odinger equation following a procedure similar to what was discussed for Eq.~(\ref{eq:jas}), except in this case we substitute $\lambda$ with the two-body binding energy.
Instead,
in the BCS case for $1/ak_F\leq -1$, the BCS wave function $\Phi_{BCS}$
gives almost the same energy as $\Phi_{FG}$ as the pairing becomes less
and less important.

Polarized systems can be simulated by extending the wave
function to include single-particle states for the unpaired
particles~\cite{Carlson_Chang_Pandharipande_etal_2003,Chang_Pandharipande_Carlson_etal_2004}. Note that the wave function as previously described cannot reproduce more exotic phases like the LOFF phase or
$p$-wave pairing.

The ground state of the system is then solved by projecting the wave
function in imaginary time by means of the DMC technique (see, for
example, Ref.~\cite{Chang_Pandharipande_Carlson_etal_2004}).  It is important to note that the DMC
algorithm strictly provides an upper bound to the energy, and thus it
is a variational calculation within the fixed-node approximation used
to control the fermion sign problem. However, if the global structure
of the wave function is carefully optimized, the DMC provides a very
accurate estimate of the ground state.  The accuracy of the DMC has been
tested by comparing the results of the equal-mass unitary Fermi gas with the
auxiliary field quantum Monte Carlo (AFQMC)~\cite{Carlson_Gandolfi_Schmidt_etal_2011}. For purely attractive
interactions, in AFQMC the unpolarized system is solved exactly
with no sign problem, although AFQMC calculations do need to account for lattice discretization effects \cite{Curry_Dissanayake_Gandolfi_etal_2024}. The agreement
between DMC and AFQMC is within 4\%-5\% giving us the confidence that
once the wave function is carefully optimized, the upper bound to the
energy given by DMC will be very close to the energy of the ground state.
It is straightforward to extend the DMC to deal with systems with
unequal masses or with polarization by modifying the wave function,
and the systematic error given by the fixed-node approximation should
provide the same accuracy as for the unpolarized case if the trial
wave function correctly describes the phase of the system. Note that for
polarized systems or for unpolarized systems with unequal masses  the
AFQMC suffers from a severe sign problem, and the constrained path used to
control it does not provide an upper bound to the energy as the fixed-node
approximation does for DMC. It should also be noted that all of the uncertainties presented on the calculations in this work are statistical in nature, arising from observables in QMC being computed as averages over many imaginary-time propagated copies of the system under study.

\section{BCS-BEC crossover}

We simulate a system of 66 fermions, as this has been shown to provide a very accurate description of fermionic systems in the thermodynamic limit~\cite{Forbes_Gandolfi_Gezerlis_2011,Gandolfi_Schmidt_Carlson_2011,Palkanoglou_Diakonos_Gezerlis_2020}. In addition it has been shown \cite{Forbes_Gandolfi_Gezerlis_2012} that near unitarity finite-size effects are minimal once the number of particles is $\geq 40$. However, since we are investigating the full crossover region, the choice of 66 particles also has the benefit of minimizing any shell effects in the weakly interaction regime, as 66 particles are a closed-shell configuration in our wavefunction. A final benefit of choosing this particular number is that results found using a BCS-like wave function can be directly
compared to the system in the normal phase. In the $\Phi_{FG}$ wave
function the single-particle orbitals are plane waves, and the energy of
33 free fermions is very close to the infinite limit.

\begin{figure}[t]
\begin{center}
\includegraphics[width=0.45\textwidth]{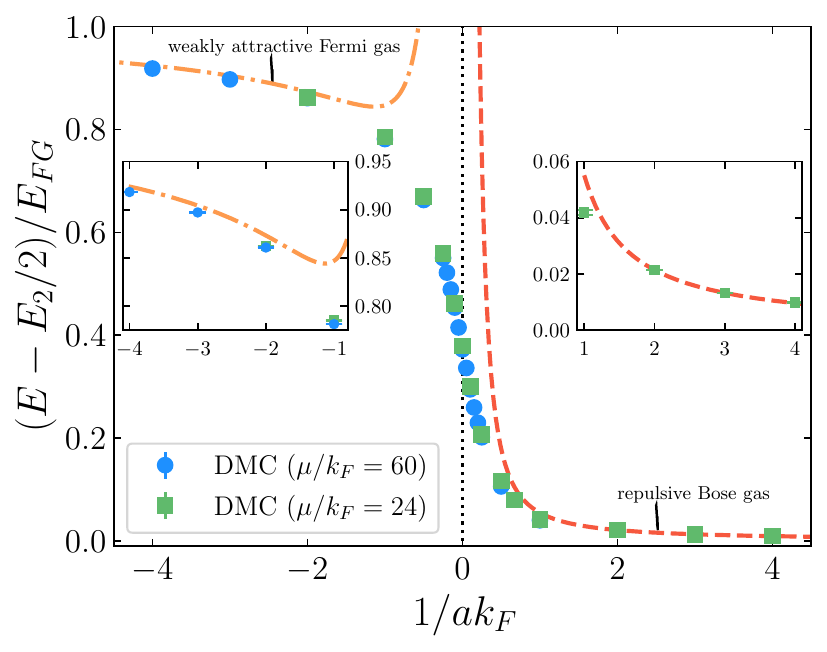}
\caption{(color online) The BCS-BEC crossover for the unpolarized
$^6$Li-$^{40}$K mixture as a function of $1/a\,k_F$. QMC results are
shown for two potentials with different effective ranges (blue circles
for $\mu/k_F=60$ and green squares for $\mu/k_F=24$).  The energy is in units of
the Fermi gas energy $E_{FG}$ after the subtraction of the two-body
binding energy per particle on the BEC side.  The orange dot-dashed line
shows the perturbation expansion of a weakly attractive Fermi gas from Eq.~(\ref{weakly})},
and the red dashed line shows the equation of state of a repulsive Bose gas from Eq.~(\ref{eq:bosegas}), with a dimer-dimer scattering length $a_{dd}=0.886(4)a$.
\label{fig:crossover}
\end{center}
\end{figure}

We show in Fig.~\ref{fig:crossover} the equation of state of the
unpolarized $^6$Li-$^{40}$K mixture as a function of $1/ak_F$. For
negative values of $a\,k_F$ the two-body interaction does not admit a
two-body bound state, and for positive values the two-body binding energy
per particle is subtracted from the total energy of the system.
The results are in units of the non-interacting Fermi gas, $E_{FG}=3E_F/5$, where $E_F=\frac{\hbar^2}{4m_r}k_F^2$.  It is clear that on the BCS side the energy per particle approaches the limit of the
noninteracting Fermi gas, which can additionally be seen by comparison with the perturbative expansion of a weakly interacting Fermi gas \cite{Huang_Yang_1957} given by, 
\begin{equation} \label{weakly}
\frac{E}{E_{FG}} = 1 + \frac{10}{9\pi}ak_F + \frac{4}{21 \pi^2}(11-2\ln 2)(ak_F)^2 + \cdots .
\end{equation}

On the BEC side, the energy per particle
quickly becomes comparable to the two-particle binding energy showing
the formation of strongly bound molecules, and $(E-E_2/2)/E_{FG}$
rapidly tends to zero.  All the points have been computed by setting
the effective range of interaction corresponding to $\mu/k_F=60$ and
tuning $v_0$ to have the desired scattering length (blue circles in
Fig.~\ref{fig:crossover}). For many points we investigated the effect of a
finite effective range and repeated the calculation using $\mu/k_F=24$
(green squares). We note that the results become independent of the effective range for $1/|a\,k_F|\ge1$, as seen by the essentially constant energy in the deep BEC regime. 

As discussed above for $1/ak_F>1$, we found that the best variational wave function is
provided by assuming the pairing wave function is the two-body solution
as in Ref.~\cite{Astrakharchik_Boronat_Casulleras_etal_2004}. By analyzing the structure of the
wave function we can conclude that for $1/ak_F\geq2$ the system made
of attractive fermions is very well modeled by a weakly repulsive Bose
gas. In this regime, if we use the pairing wave function in Eq.~(\ref{eq:pairfn}),
the energy of the system is higher, and we would likely need many
more plane waves to correctly describe the short-range behavior of the
two-particle solution.

\begin{figure}[t]
\begin{center}
\includegraphics[width=0.45\textwidth]{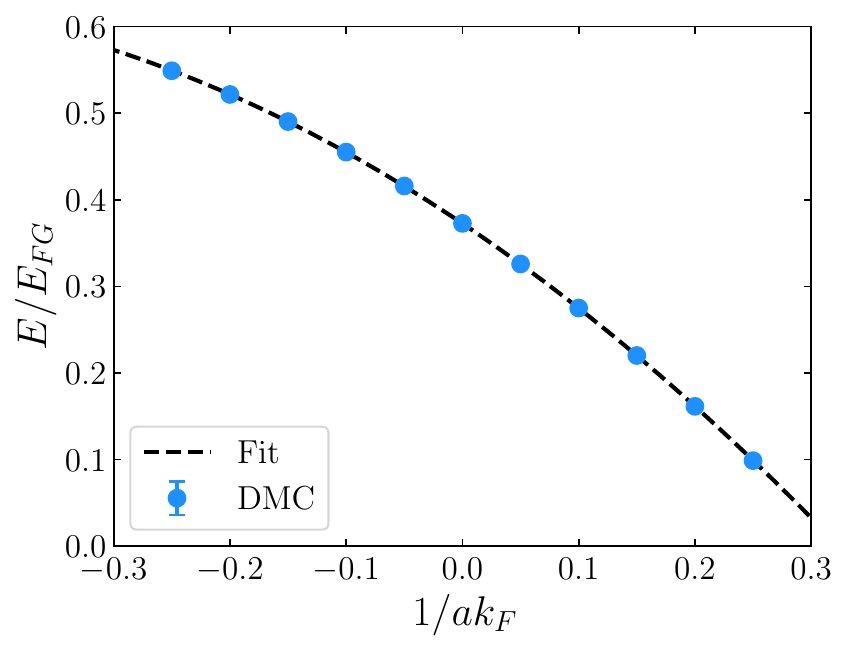}
\caption{(color online) The BCS-BEC crossover for the unpolarized
$^6$Li-$^{40}$K mixture as a function of $1/a\,k_F$ near the unitary limit. Our results are compared to a fit given by Eq.~(\ref{eq:eos}) as discussed in the text.}
\label{fig:nearunitarity}
\end{center}
\end{figure}

The energy of a repulsive Bose gas, in which bosons are made by two
fermions with different masses, can be well approximated by a mean-field
expansion~\cite{Lee_Huang_Yang_1957},which for unequal masses is given by
\begin{eqnarray}
\label{eq:bosegas}
\frac{E/N-E_2/2}{E_{FG}}&=&
\frac{10}{9\pi}a_{dd}k_F\frac{m_h m_l}{(m_h+m_l)^2}
\nonumber \\
&& \times \left[1+\frac{128}{15\sqrt{6\pi^3}}(a_{dd}k_F)^{3/2}+\dots\right],
\end{eqnarray}
where $a_{dd}$ is the boson-boson scattering length, whose value
is obtained by fitting to our QMC calculations. The results in the strongly
interacting regime are shown in the right inset of
Fig.~\ref{fig:crossover} for the mass ratio considered. The fit is made in the $1/ak_F\geq 2$ region, giving $a_{dd}/a=0.886(4)$,
in good agreement with the few-body analysis of
Refs.~\cite{Petrov_Salomon_Shlyapnikov_2005,Levinsen_Petrov_2011}.  The agreement between the QMC
results and the mean field expansion of Eq.~(\ref{eq:bosegas}) starts to
fail for $1/ak_F<2$, indicating that the system is no longer similar
to a Bose gas, as also suggested by looking at the structure of the many-body
wave function.

In Fig.~\ref{fig:nearunitarity} we show the crossover
around the unitary limit.  Close to unitarity, the equation of
state can be expanded as~\cite{Tan_2008b}
\begin{equation}
\label{eq:eos}
\frac{E}{E_{FG}} = \xi-\frac{\zeta}{k_F a}-
\frac{5\nu}{3(k_F a)^2}+\dots \,.
\end{equation}

\begin{figure}[t]
\begin{center}
\includegraphics[width=0.48\textwidth]{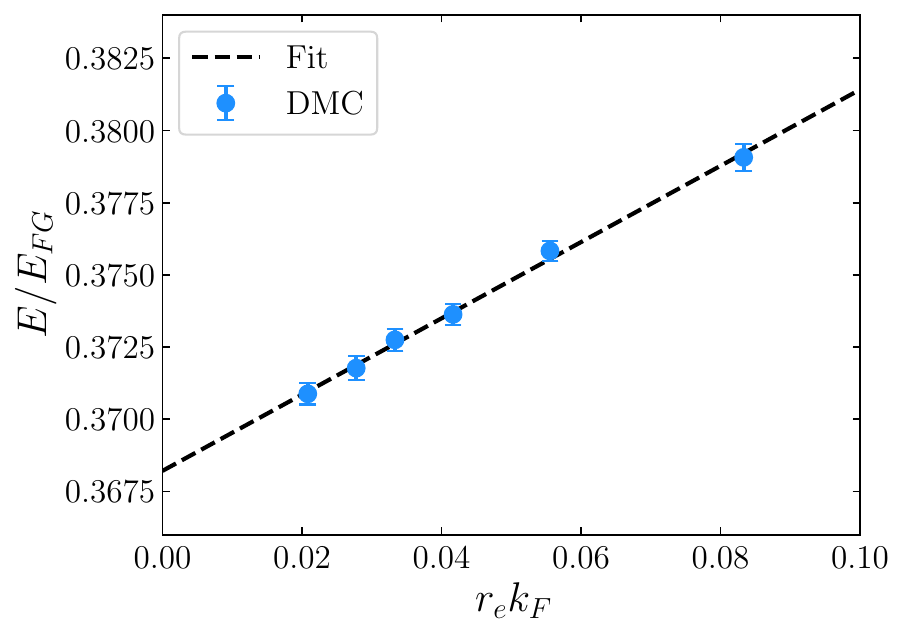}
\caption{(color online) The energy of the $^6$Li-$^{40}$K mixture as a
function of the effective range $r_{e}$ of the interaction in the
unitary limit. The energy is in units of the Fermi gas energy $E_{FG}$.
The slope of the extrapolation is similar to what was found in the equal-mass case with lattice calculations \cite{Carlson_Gandolfi_Schmidt_etal_2011}.
The details of the fit to the QMC results are discussed in the text.}
\label{fig:reff}
\end{center}
\end{figure}

The fit to our QMC results near unitarity gives ${\xi=0.3726(6)}$, ${\zeta=0.900(2)}$, and
${\nu=0.46(1)}$.  Our result for $\xi$ differs appreciably from the equal-mass case near unitarity ($\xi=0.383(1)$ from \cite{Gandolfi_Schmidt_Carlson_2011}), which is expected to be a result of non-zero total momentum pairs in the ground state \cite{Gezerlis_Gandolfi_Schmidt_etal_2009}. However, the values of $\zeta$ and $\nu$ are very similar to what was found for the equal-mass case Ref.~\cite{Gandolfi_Schmidt_Carlson_2011}. Our results are in good qualitative agreement with the calculation of
Ref.~\cite{Diener_Randeria_2010}, deviating from mean-field calculations which predict no difference between the equal- and unequal-mass cases.

The various tests we performed indicate that around ${1/a\,k_F\approx0}$
the extrapolation to $r_{e}\rightarrow 0$ just shifts the energy and
does not change the curvature of the equation of state in the region
where $1/|a\,k_F|<1$. There, only the $\xi$ parameter is particularly
sensitive to $r_{e}$, which is also the case for the equal-mass system \cite{Gandolfi_Schmidt_Carlson_2011}.
It is interesting to compare the finite-range extrapolation for the
unequal-mass mixture to the equal-mass case. The comparison is
warranted for the $^6$Li-$^{40}$K system because the resonance is narrower than
for equal masses~\cite{Naik_Trenkwalder_Kohstall_etal_2011}, and consequently, the unitary limit
could be ``less universal'', and the effect of the effective range could be more
important.  We performed several calculations at unitarity for different
values of $r_{e}$ in order to test the finite-range effects and extrapolated to zero effective range as
shown in Fig.~\ref{fig:reff}.  The QMC results were fit using the
linear function 
\begin{equation} 
\frac{E}{E_{FG}}=\xi+c\,r_{e}k_F \,,
\end{equation} 
and the best estimate given by QMC is $\xi=0.368(1)$. Similar to the above case near unitarity, our value for $\xi$ at the unitarity limit is slightly lower than what is found in equal-mass calculations carried out on the lattice ($\xi=0.372(3)$ from \cite{Carlson_Gandolfi_Schmidt_etal_2011}. This is further evidence of the contribution of nonzero total momentum pairs in the ground state, as found in \cite{Gezerlis_Gandolfi_Schmidt_etal_2009}. The value of $c=0.13(1)$ gives the slope of the extrapolation, and it is rather intriguing that the slope of the energy as a function of the effective
range is very similar to the equal-mass case~\cite{Carlson_Gandolfi_Schmidt_etal_2011}. This fact, along with the similarities to higher order terms from Eq.~(\ref{eq:eos}), should be explored in future investigations into mass-imbalanced cold atomic systems.

As discussed above, there are many reasons to expect different physics between the unequal-mass Fermi gas that we study and the more common equal-mass case. At the level of the equation of state the two systems display certain similarities (higher-order terms in the equation of state), as well as differences (the Bertsch parameter $\xi$ is quantitatively different near and at the unitary limit). To investigate further similarities and differences between the equal- and unequal-mass Fermi gases, we consider a number of different physical observables.

\begin{figure}[t]
\begin{center}
\includegraphics[width=0.45\textwidth]{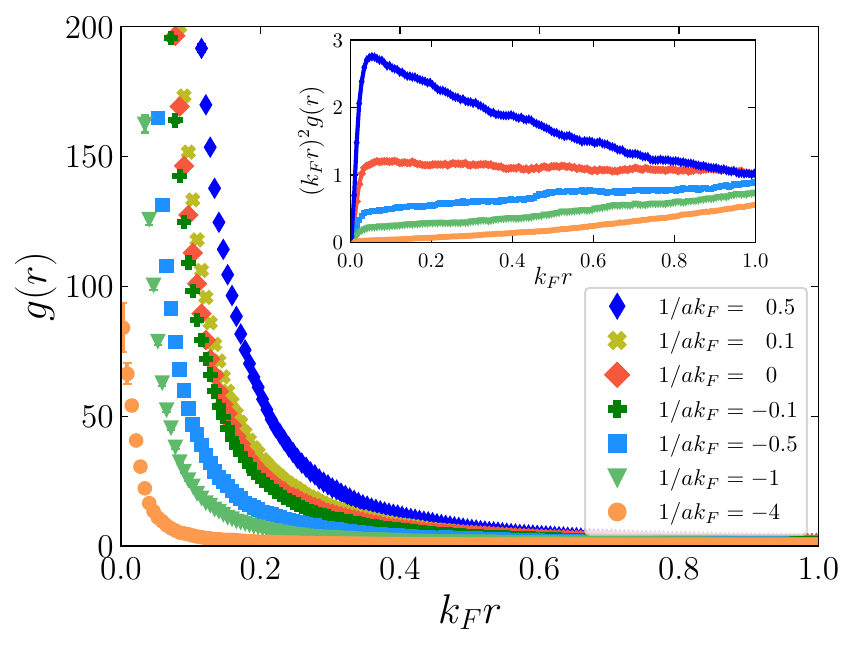}
\caption{(color online) Heavy-light pair-distribution function for
different two-body interactions indicated in the legend. From left to
right they correspond to $1/ak_F=-4$, $-1$, $-0.5$, $-0.1$, $0$, $0.1$, and
$0.5$.  In the inset we show the same function multiplied by $(k_Fr)^2$
for selected interaction strengths corresponding to $1/ak_F=-4$, $-1$,
$-0.5$, $0$, and $0.5$ (from the bottom to the top).
}
\label{fig:gofr}
\end{center}
\end{figure}

We first consider the pair-distribution function $g(r)$ between heavy
and light particles, as shown in Fig.~\ref{fig:gofr}. $g(r)$ is typically calculated by a mixed estimate of the form
\begin{align}
    g(r) = \sum_{i<j'} \frac{\bra{\Psi_0}\delta (r_{ij'}-r)O_{ij'}^P \ket{\Psi_T}}{\braket{\Psi_0|\Psi_T}},
\end{align}
where $\Psi_0$ is the DMC projected ground state and, for our purposes ($s$-wave central interaction) the $O_{ij'}^p$ operator is equal to 1 \cite{Gezerlis_Carlson_2010}. The sum is only over pairs of particles with different masses.

The function
$g(r)$ exhibits a large peak at short distances, induced by the
strong attractive interaction in the $s$-wave channel. In the weak-coupling limit the
value of the peak is smaller as expected, and is clear in our results.
In the inset of Fig.~\ref{fig:gofr} we show $(k_Fr)^2 g(r)$, which
can be used to extract the contact parameter, as shown in
Ref.~\cite{Gandolfi_Schmidt_Carlson_2011}.
The limit of $(k_Fr)^2
g(r)$ for short distances approaches zero in the BCS limit
as the short-range correlations are weaker when the strength of the
attractive interaction decreases. On the other hand, in the strong-coupling regime where $1/ak_F=0.5$, the limit of $(k_Fr)^2 g(r)$ at
small distances rapidly increases. Our results for unitarity (red diamonds in Fig.~\ref{fig:gofr}) can be compared with the equal-mass case shown in Fig.~4 of \cite{Gandolfi_Schmidt_Carlson_2011}. One notable difference is that in the equal-mass case $g(r)$ decays with increasing $k_F r$, and we find this effect to be dramatically reduced for the unequal-mass case. By looking at the inset of Fig.~\ref{fig:gofr}, at unitarity we see that $(k_Fr)^2g(r)$ is essentially constant over the considered range of $k_Fr$. This can be interpreted as the heavy-light pairs remaining farther apart from one another, potentially due to the higher inertia of the heavy particles, compared to the equal-mass case.

\begin{figure}[t]
\begin{center}
\includegraphics[width=0.45\textwidth]{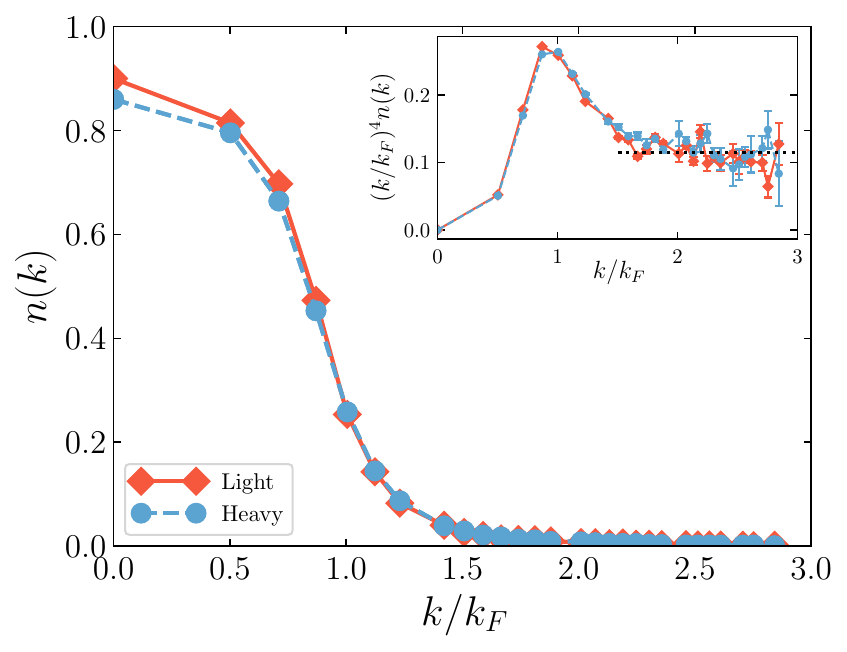}
\caption{(color online) Momentum distribution $n(k)$ for the unequal-mass system at unitarity. We calculate $n(k)$ using only the light particles (solid line) and using only the heavy particles (dashed line). In the inset we plot $(k/k_F)^4n(k)$ to show the tail behavior of the momentum distribution. The dotted black line shows the contact parameter $C$ derived from our calculated equation of state around the unitary limit, following the prescription from \cite{Werner_Castin_2012}}.

\label{fig:nofk_unitarity}
\end{center}
\end{figure}

\begin{figure}[t]
\begin{center}
\includegraphics[width=0.45\textwidth]{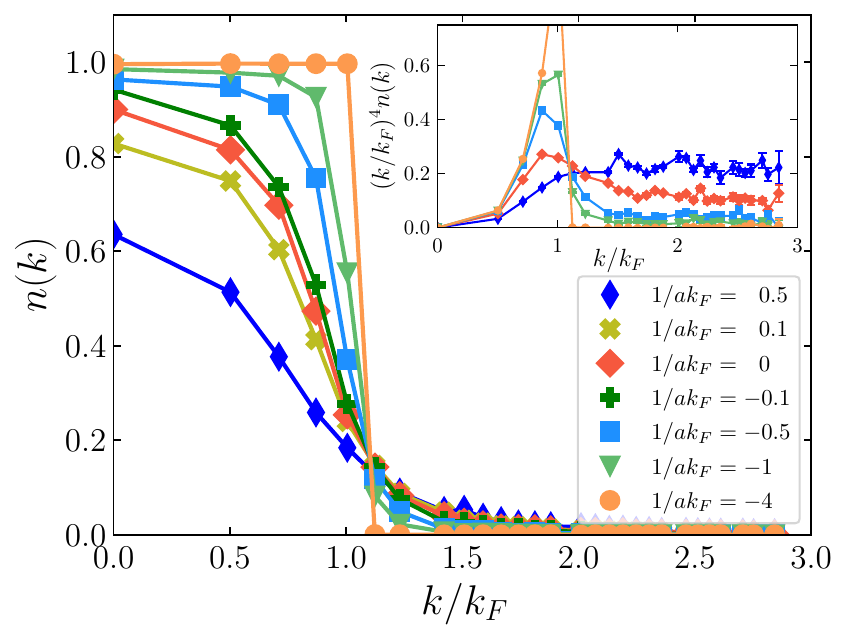}
\caption{(color online) Momentum distribution $n(k)$ for different
interaction strengths (same color convention as in Fig.~\ref{fig:gofr}), using only the light particles.
In the inset we plot $(k/k_F)^4n(k)$ to show the tail behavior of
the momentum distribution. The results for unitarity (red diamonds) also appear in Fig.~\ref{fig:nofk_unitarity}.}
\label{fig:nofk}
\end{center}
\end{figure}

In Figs. \ref{fig:nofk_unitarity} and \ref{fig:nofk} we show momentum distributions for the mass imbalanced system, calculated as the Fourier transform of the one-body density matrix by,
\begin{align} \label{nofk}
    n(k) = \frac{N}{L^3} \hspace{-0.1cm} \int \hspace{-0.1cm} d\Omega d \delta r e^{i\bm{k}\cdot (\bm{r}'_N - \bm{r}_N)}\frac{\Psi_T(\bm{r}_1,\cdots,\bm{r}'_{N})}{\Psi_T(\bm{r}_1,\cdots,\bm{r}_{N})} ,
\end{align}
where the integral over the solid angle is handled stochastically and the integral over $\delta r = |\bm{r}'_n - \bm{r}_n|$ is carried out on a line to avoid statistical errors \cite{Gezerlis_Carlson_2010}. The primed coordinate refers to a single particle that is moved along this line in order to carry out the integration over $d\delta r$. Equation (\ref{nofk}) can then be computed for each particle in our system, and an average value can be taken. In Fig.~\ref{fig:nofk_unitarity} we first investigate the momentum distribution for the system at unitarity, averaging over only the light or heavy particles independently. We also investigate the tail behavior of the momentum distribution by plotting $(k/k_F)^4 n(k)$. We find in both cases a value for Tan's contact parameter that agrees with the expected result from \cite{Werner_Castin_2012}, given our equation of state in Fig.~\ref{fig:crossover}. Figure \ref{fig:nofk_unitarity} also highlights a subtle difference between the momentum distributions of the heavy and light particles, with the heavy-particle calculations showing a noticeable decrease for low-momentum values. This feature is absent in studies of the equal-mass Fermi gas.

\begin{figure}[t]
\begin{center}
\includegraphics[width=0.45\textwidth]{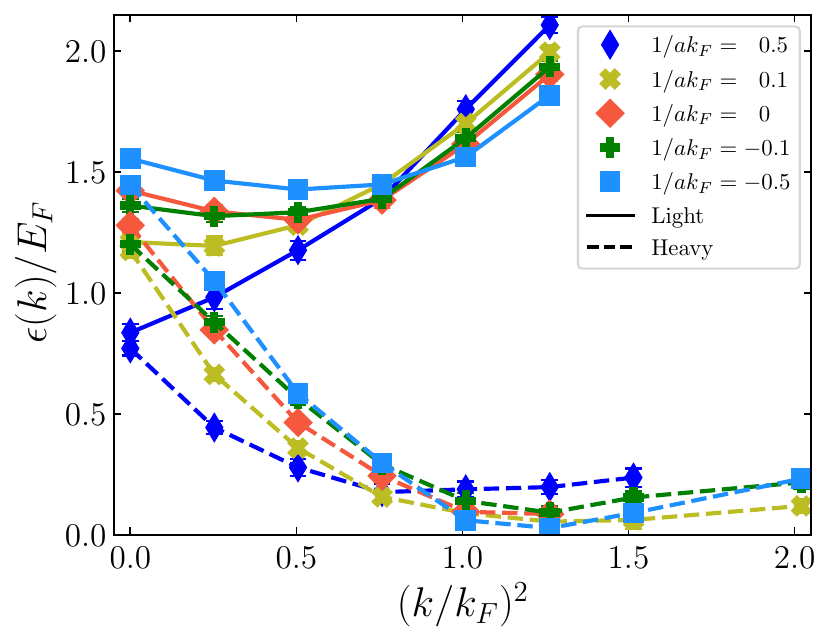}
\caption{(color online) Quasiparticle dispersion at different
scattering lengths.  The upper and lower curves show results for light
(solid lines) and heavy (dashed lines) quasiparticles as a function
of the momentum. We show results for the unitary limit (red diamonds) as well as the BCS region ($ak_F < 0$) and the strong interacting BEC region ($ak_F > 0)$. All results correspond to a mass ratio $m_h/m_l=6.5$.}
\label{fig:quasipart}
\end{center}
\end{figure}

Since considering either the light or heavy particles alone gives the same value for the contact, we can investigate the crossover behavior of the momentum distribution, as shown in Fig.~\ref{fig:nofk}, considering only the light particles. In the very weak coupling region with $1/ak_F=-4$, $n(k)$ is very similar to
the non interacting Fermi gas momentum distribution, where $n(k)=1$
for $k<k_F$ and zero otherwise. Moving to strong couplings the tail
of $n(k)$ is longer, as expected, due to the strong pairing effects. This is
clearly evident in the inset of Fig.~\ref{fig:nofk} where the function
$(k/k_F)^4 n(k)$ is shown.  The inset clearly shows the $C/k^4$
universal behavior predicted by Tan \cite{Tan_2008,Tan_2008b,Tan_2008c}, where $C$ is the contact, regardless of interaction strength. The value of $C$ we find from the tail of $n(k)$ agrees with the value that can be computed from a derivative of the equation of state with respect to the scattering length \cite{Werner_Castin_2012}. It is also interesting to note that the value of $C$ we find for the unequal-mass case agrees within error with what is found for the equal-mass case \cite{Gandolfi_Schmidt_Carlson_2011}. This highlights a fundamental similarity between the equal- and unequal-mass Fermi gases.

Last, the quasiparticle dispersion is
computed by adding an unpaired particle to the
system~\cite{Carlson_Chang_Pandharipande_etal_2003,Chang_Pandharipande_Carlson_etal_2004,Gezerlis_Gandolfi_Schmidt_etal_2009,Gandolfi_Illarionov_Fantoni_etal_2008,Gezerlis_Carlson_2010}.
We carried out simulations of 67 particles where the unpaired particle
is characterized by a finite momentum $|k|$ and mass $m_l$ or $m_h$.
We then combine the result with the energy of the fully paired unpolarized
system of 66 particles and define the quasiparticle energy as
\begin{equation} 
\epsilon(k)/E_{F}=[E_k(67)-E(66)]/E_{F} \,, 
\end{equation} 
where $k$ is the momentum of the extra particle, and $E(67)$ and
$E(66)$ are the total energies of the systems with 67 and 66 particles.
The two simulations were performed at the same total density.
In Ref.~\cite{Gezerlis_Gandolfi_Schmidt_etal_2009} the quasiparticle dispersion was
computed by comparing the energy of the two systems at the same background
density, then at constant volume. Here we have fixed the total density
in order to avoid small effects due to the effective range of the
interaction. However, the finite-size effects should be very small as
the volume of the simulation box is quite large.

The results of the quasiparticle dispersion in
Fig.~\ref{fig:quasipart} show that the dispersions of light and heavy particles
are very different from what was already found at unitarity~\cite{Gezerlis_Gandolfi_Schmidt_etal_2009}.
For each case the quasiparticle dispersion is lower on the
BCS side and increases by crossing unitarity and moving to the BEC side.
In particular for light quasiparticles the position of the minimum changes
from about $(k/k_F)^2=0.5$ at $ak_F=-2$ to $(k/k_F)^2=0$ at $ak_F=2$. Our results at unitarity agree with previous findings  \cite{Gezerlis_Gandolfi_Schmidt_etal_2009} that the quasiparticle dispersion for the unequal-mass Fermi gas differs dramatically from the equal-mass case \cite{Carlson_Reddy_2005}. We have also shown that the differences in the quasiparticle dispersions for heavy and light particles are seen throughout the BCS-BEC crossover region.

\section{Conclusions}
In this paper we showed QMC results exploring the BCS-BEC
crossover of the mass-imbalanced $^6$Li-$^{40}$K Fermi mixture.  QMC techniques were used in the past to give the best upper bound of the energy of strongly
interacting fermions in the unitary limit. We made use of QMC to compute
the equation of state as a function of the two-body scattering length,
studying the crossover from the weakly interacting limit where the system
is well described by the BCS theory to the strong coupling limit where
fermions form molecules, and the system is well described by a weakly
repulsive Bose gas.  We carefully checked finite-range effects close
to the unitary limit and provided an accurate fit of the equation of
state based on our QMC calculations.  We also showed other properties of the
unpolarized superfluid system, namely, the pair-distribution function and
the momentum distribution.  Finally, by considering a system with very
small polarizations, we computed the heavy- and light-quasiparticle
dispersions for different strengths of the interaction. Throughout we also paid careful attention to the differences and similarities between the equal- and unequal-mass Fermi gases, finding several important results. Our calculations of the equation of state revealed that the higher order terms in the equation of state are very similar. We also found that the universal contact parameter is identical to that in the equal-mass case. However, we did find several key differences from the equal-mass case, such as the Bertsch parameter, quasiparticle dispersion, and tail of the pair-distribution function. Our results should provide a comprehensive framework for future theoretical and experimental investigations into mass-imbalanced Fermi gases across the full range of the BCS-BCS crossover.

\section*{Acknowledgments} 
The work of S.G. is supported by the U.S. Department of Energy,
Office of Nuclear Physics, under Contract No. DE-AC52-06NA25396, and
by the Office of Advanced Scientific Computing Research, Scientific
Discovery through Advanced Computing (SciDAC) NUCLEI program.
The work of R.C. and A.G. was supported by the Natural Sciences and Engineering Research Council (NSERC) of Canada and the
Canada Foundation for Innovation (CFI). R.C. was also supported by the Laboratory Directed Research and Development program of Los Alamos National Laboratory under Project No. 20220541ECR.
Computational resources were provided by the Los Alamos National
Laboratory Institutional Computing Program, which is supported by the U.S. Department of Energy National 
Nuclear Security Administration under Contract No. 89233218CNA000001.
Computer time was also made available by the National Energy Research Scientific Computing
Center (NERSC) and SHARCNET through the Digital Research Alliance of Canada.

\bibliographystyle{apsrev4-1}
\bibliography{biblio}

\begin{thebibliography}{79}
\bibitem [{For the case of the central interactions in this work the diffusion Monte Carlo approach is exactly equivalent to, and occasionally referred to as()}]{footnote}%
  \BibitemOpen
  For the case of the central interactions in this work the diffusion Monte Carlo approach is exactly equivalent to, and occasionally referred to as,\ \href@noop {} {}\bibinfo {note} {Green's function Monte Carlo}\BibitemShut {NoStop}%
\end{thebibliography}


\end{document}